# Frustrated Magnetism in Triangular Lattice TlYbS$_2$ Crystals Grown via Molten Flux

**Timothy Ferreira**[a*], **Jie Xing**[a], **Liurukara D. Sanjeewa**[a], **Athena S. Sefat**[a]

*Materials Science and Technology Division, Oak Ridge National Laboratory, Oak Ridge, Tennessee 37831, USA*

**\* Correspondence:**
Tim Ferreira
ferreirate@ornl.edu

**Abstract**

The triangular lattice compound TlYbS$_2$ was prepared as large single crystals via a molten flux growth technique using sodium chloride. Anisotropic magnetic susceptibility measurements down to 0.4 K indicate a complete absence of long-range magnetic order. Despite this lack of long-range order, short-range antiferromagnetic interactions are evidenced through broad transitions, suggesting frustrated behavior. Variable magnetic field measurements reveal metamagnetic behavior at temperatures ≤ 2 K. Complex low temperature field-tunable magnetic behavior, in addition to no observable long-range order down to 0.4 K, suggest that TlYbS$_2$ is a frustrated magnet and a possible quantum spin liquid candidate.

**INTRODUCTION**

The variety of basic and applied properties arising from geometrically frustrated magnets continues to motivate the study of structures with two-dimensionally layered triangular lattices prone to antiferromagnetic interactions.[1–7] Such triangular lattices restrict the number of available spin degrees of freedom, resulting in quantum fluctuations that can produce degenerate ground states.[8] Frustrated antiferromagnets with degenerate ground states have garnered significant interest for their potential as quantum spin liquid (QSL) candidates, a state characterized by dynamic entangled spins, exhibiting no long-range magnetic order, even at 0 K.[5, 8–18] While QSL candidates with 3$d$ ions exhibit weak spin-orbit coupling (SOC),[19–24] the presence of stronger SOC in 5$d$/4$f$ ions, on par with the energy scale of crystal electric field effects and the coulomb interaction $U$, further enhances the frustration via entangled spin and orbital degrees of freedom, and has thus shifted the search for new QSL materials to contain these heavier lanthanides (*Ln*).[16, 25–33]

Recently, Yb(III) containing compounds such as NaYbO$_2$,[1, 2] NaYbS$_2$,[9] and YbMgGaO$_4$[4, 10, 11, 13, 34] have been presented as QSL candidates, all crystallizing in a layered triangular lattice of trigonal space group *R*-3*m*. Due to the odd number of 4$f$ electrons and strong SOC, these materials behave as effective spin $J_{\text{eff}} = ½$. A similar family of 4$f$-containing delafossites, of the general formula A(I)*Ln*(III)*Ch*(II) [A = Na, Rb, K; *Ch* = O, S, Se, Te],[31] has also been proposed as a promising candidate. Delafossites are often free from crystallographic site-mixing, unlike the more commonly studied YbMgGaO$_4$ that can mimic QSL behavior by eliminating long-range order through disorder.[10,35] Additionally, the modular nature of delafossite structures allows for the possibility of differences in crystal structure as a function of the ratio of ionic radii; this has been reported to result in changes in triangular lattice layer stacking, such as ABAB stacking in the hexagonal *P*6$_3$/*mmc* or ABCABC layer stacking in the trigonal *R*$\bar{3}$*m*. Reports of possible inter-layer interactions on the highly sensitive magnetic ground state of such systems makes the delafossite structure advantageous to study, as such subtle interactions can be probed as a function of selective ion control.[18]

Despite the structural modularity and promise of QSL candidacy in such delafossite structures, the limited availability of large single crystals to study electronic and magnetic anisotropy serves as motivation for this work. Herein we discuss the flux crystal growth, structure determination, and magnetic property measurements of TlYbS$_2$, which crystallizes in the hexagonal space group *P*6$_3$/*mmc*. This study allows for proper structure elucidation of TlYbS$_2$ single crystal contrary to the results of polycrystalline powders,[36] in addition to reporting novel anisotropic magnetization results that were inaccessible on polycrystalline samples.[36]

**EXPERIMENTAL**

**Synthesis**

The TlYbS$_2$ compound was synthesized using a two-step method comprised of (1) producing the powder form via traditional solid-state synthesis, followed by (2) crystallizing the precursor powder using molten flux growth via sodium chloride. Solid pieces of Yb metal (REacton, 99.99%), S (Puratronic, 99.9995%) and Tl (REacton, 99.99%) were all stored in a glove box. Sodium chloride (Alfa Aesar, 99.999%) was dried in an oven overnight at 300 °C and stored in a desiccator prior to use.

For the solid-state synthesis, 2.0 mmol of Tl and Yb were added to 4.0 mmol of S in an alumina crucible with a loose-fitting alumina frit and a second, inverted, alumina crucible on top. This second crucible was used to assist in catching any trace amounts of volatilized Tl or Yb metals. This setup of alumina crucibles was loaded and sealed inside an evacuated silica tubing, with a small amount of quartz wool at the bottom to prevent cracking due to differences in thermal expansion. (**Figure 1a**) The sealed silica tube was heated to 300 °C at a rate of 10 °C/hr, dwelled for 24 hr, ramped to 800 °C at a rate of 10 °C/hr, dwelled at 72 hr, and then the reaction was allowed to cool by shutting off the furnace.

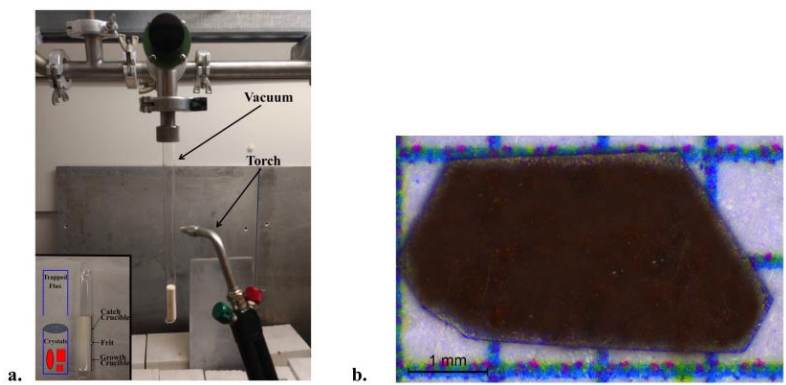

Figure 1. (a) Experimental setup consists of an alumina crucible housing the starting materials with a loose-fitting frit and a second inverted alumina crucible, shown in the inset, both held within a sealed evacuated quartz tube. (b) Optical image of a typical piece of TlYbS$_2$, showing its red color. The *c* axis is out of the plane of the crystal.

Single crystals were produced by loading 0.87 mmol of polycrystalline $TlYbS_2$ and a ten-fold excess (by mass) of NaCl (40.47 mmol) into sealed evacuated silica tubing. The reaction was heated to 850 °C at a rate of 60 °C/hr, dwelled at 504 hr (i.e., three weeks), and then cooled by shutting off the furnace. The resulting transparent red crystals were mechanically separated from the remaining $TlYbS_2$ powder and vacuum filtered using ethanol to remove any surface impurities. **(Figure 1b)** The crystallographic *c*-axis is out of the plane of the paper. The purity of the polycrystalline powder (first step) and resulting crystals (second step) were both determined by powder X-ray diffraction (PXRD) using a PANalytical X'Pert Pro MPD diffractometer with Cu Kα1 radiation (λ = 1.5418 Å), shown in **Figure 2**.

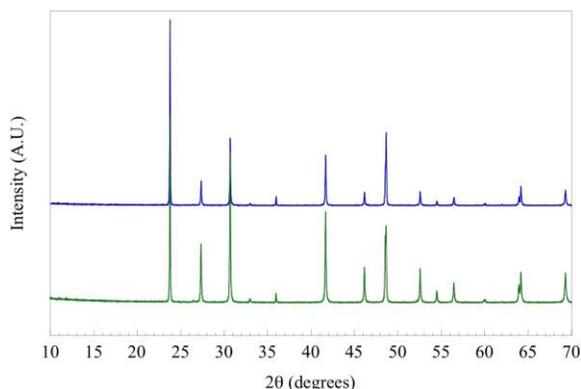

Figure 2. Powder X-ray diffraction pattern of solid state-produced polycrystalline $TlYbS_2$, shown in blue, overlaid with a diffraction pattern of ground single crystals of $TlYbS_2$, shown in green.

**Structure**

The structure determination of the single crystals of $TlYbS_2$ was performed on a Bruker Quest D8 single crystal X-ray diffractometer (SXRD). Single crystals were first sonicated in acetone to remove any surface impurities. The data were collected at room temperature utilizing a Mo Kα radiation, (λ = 0.71073 Å). The crystal diffraction images were collected using Φ and ω-scans. The diffractometer was equipped with an Incoatec IµS source using the APEXIII software suite for data setup, collection, and processing.[37] The structure was resolved using intrinsic phasing and full-matrix least square methods with refinement on $F^2$ using the SHELXTL software suite.[38] All atoms were first refined with isotropic thermal displacement parameters and then refined anisotropically. Crystallographic information can be found in **Table 1** to **3**.

Energy dispersive spectroscopy (EDS) was performed using a Hitachi S-3400 scanning electron microscope equipped with an OXFORD EDX microprobe to confirm the elemental composition in the single crystal sample. The presence of Tl, Yb and S were verified, and the absence of extraneous elements such as Na and Cl were confirmed. EDS data can be found in **Table 4**.

**Magnetic Susceptibility**

Physical properties were measured using the Quantum Design Superconducting quantum interference device (SQUID) Magnetic Properties Measurement System (MPMS). Two thin plates of crystals were stacked with a total mass of 0.36 mg using vacuum grease, for each measurement above 2 K. The magnetization measurements made below 2 K were measured using the Quantum Design iHe3 option on four stacked single crystals (0.5 mg total) using vacuum grease. The crystals were aligned in two separate orientations: one set of measurements aligned the crystals such that the applied magnetic field was perpendicular to the *c*-axis, and second set where the applied field was parallel with *c*. The thermometer calibration was done in zero field when the applied field was parallel with c.

**DISCUSSION**

**Synthesis**

The scarcity of known single crystal growth methods for lanthanide containing delafossites, particularly those employing the use of salt flux, has made the optimization and characterization of large single crystals of the titled composition challenging.[39] Fortunately, there is significant literature evidence demonstrating that the use of molten flux as a growth medium is a robust method, potentially capable of crystallizing nearly every element combination on the periodic table.[40] The selection of an alkali halide flux was guided in part by recent reports of similar fluxes crystallizing compositions containing lanthanides and/or chalcogenides.[41–44] Additionally, amongst the few delafossite-type structures reported as single crystals, synthesis typically involves the use of reactive alkali fluxes, such that the alkali metal in the flux incorporates into the final product, such as the use of KCl for $KErSe_2$[45] or NaCl for $NaYbS_2$.[9] For producing $TlYbS_2$ crystals, the use of TlCl as a flux medium was ruled out due to the low solubility prohibiting ease of crystal separation upon completion of the reaction. We tried $CaCl_2$ as a flux, and although single crystals were produced, they were of poor quality. Attempts to improve the quality of the crystals by introducing cooling rates also failed: a variety of polycrystalline powders and no crystals were present, indicating $TlYbS_2$ may be a metastable kinetic phase that is 'trapped' via quenching. Ultimately, the use of NaCl as a flux, in addition to quenching the reaction upon the completion of the prolonged dwelling period, resulted in high quality, large (1 mm+) single crystals that were suitable for structural and anisotropic magnetization studies.

**Structure**

The compound $TlYbS_2$ was first reported as a polycrystalline powder crystallizing in the trigonal space group *R-3m*, commonly referred to as the α-$NaFeO_2$ structure, with lattice parameters $a = 3.935$ Å and $c = 22.47$ Å.[36] This two dimensional layered structure is built from two distinct triangular lattice layers that alternate along the *c* axis. The first layer is built from a network of edge-shared octahedral $YbS_6$ units that adopts an ABCABC stacking pattern. (**Figure 3**) A second non-magnetic triangular lattice layer of edge-shared octahedral $TlS_6$ units resides between these layers. As a result of the availability of single crystals of the titled composition, we thoroughly investigated the nuclear structure. Careful analysis revealed that the grown single crystals of $TlYbS_2$ adopt the hexagonal *P*6$_3$/*mmc* β-$RbScO_2$ structure type. In this hexagonal structure, Tl (Wyckoff 2*c*), Yb (2*b*), and S (4*f*) occupy the special positions with site symmetries of -6*m*2, -3*m*, and 3*m*, respectively. This is in contrast to the previously reported trigonal *R-3m* structure for polycrystalline powders of $TlYbS_2$ where Tl (3*b*) and Yb (3*a*) are in the -3*m* position, and S (6*c*) is in the 3*m* special position. The primary difference between the two structures is best understood by the number and stacking sequence of the triangular lattice layers. In the hexagonal

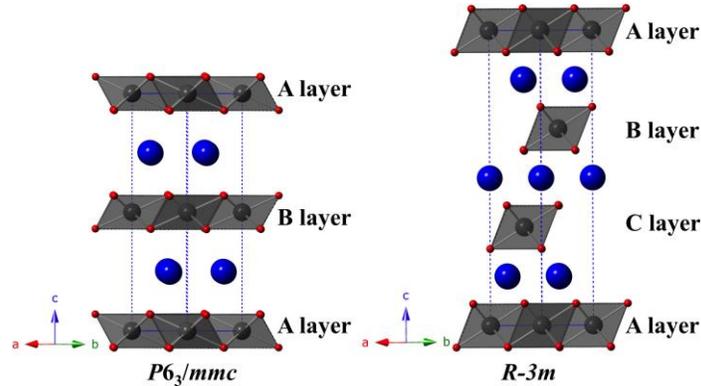

Figure 3. Polyhedral representation of the delafossite structure, with the hexagonal *P*6$_3$/*mmc* variant (shown left) and the trigonal *R*-3*m* variant (shown right). The hexagonal structure is built of three distinct triangular layers of edge-shared YbS$_6$ octahedra (Yb shown in grey, S in red) that stack in an ABAB order, meanwhile the trigonal structure is built from four distinct triangular layers of edge-shared YbS$_6$ octahedra that stack in an ABCABC sequence. Non-magnetic Tl atoms are shown as blue spheres.

β-RbScO$_2$ structure, one unit cell is built from three YbS$_6$ layers and two TlS$_6$ layers, whereas the trigonal α-NaFeO$_2$ structure unit cell is built from four YbS$_6$ layers and three TlS$_6$ layers. Additionally, the hexagonal β-RbScO$_2$ structure adopts a higher symmetry ABAB triangular lattice layer stacking, in contrast to the ABCABC layer stacking in the trigonal α-NaFeO$_2$ type structure. This result is consistent with the trend observed for smaller A-site ions in the delafossite structure crystallizing in the trigonal system, such as NaYbS$_2$,[9] and larger A-site ions crystallizing in the hexagonal system, such as CsYbSe$_2$.[46] Comparison of powder X-ray diffraction patterns of solid-state produced polycrystalline powder, and single crystals grown via molten flux, of TlYbS$_2$ overlay well (**Figure 2**) with slight differences in peak intensity and crystallinity. Although polymorphism between the polycrystals and single crystals should not be completely ruled out, slight differences in peak intensity may result from preferred orientation in such a highly anisotropic nuclear structure.

The primary interest in studying this structure type is the triangular lattice that extends along the *ab* plane, serving as an ideal host for geometric frustration. (**Figure 4**)

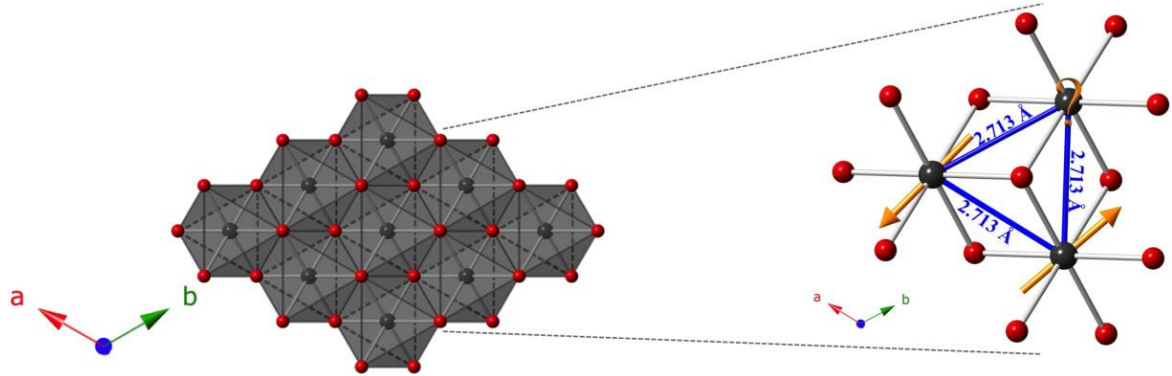

Figure 4. Polyhedral representation of the edge-shared octahedral network of YbS$_6$ extending in the *ab* plane for the composition TlYbS$_2$. The planar triangular lattice is an ideal host for geometric frustration for an antiparallel alignment of spins, as shown in the zoomed-in area. Yb atoms are shown in grey, and S atoms are shown in red.

The idealized hexagonal network of Yb atoms does not allow for a purely antiparallel configuration of spins, leading to enhanced quantum fluctuations that are of interest to study. Careful attention was taken to select Tl as the A site in the delafossite structure, since it is larger than Yb, both to prevent crystallographic site-mixing, and to maximize the interlayer distances, thereby minimizing inter-layer interactions via the mediating cation. The reactive-flux nature of NaCl in the growth of similar delafossite structures also guided the selection of the large cation Tl, in hopes that Na would be too small to occupy the same site. Additionally, a non-magnetic A site was selected to further simplify study of any resulting magnetic properties.

**Magnetic Susceptibility**

The temperature dependence of the magnetization perpendicular to the applied field (in the *ab* plane) at 1 T indicates a broad deviation from Curie-Weiss behavior at 50 K. (**Figure 5**)

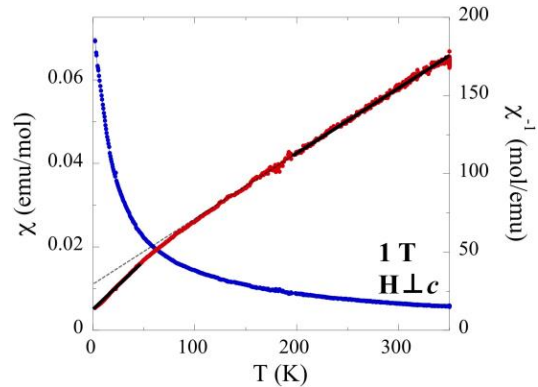

Figure 5. Temperature dependence of the magnetic susceptibility for single crystals of TlYbS$_2$ oriented H$\perp c$ under an applied field of 1 T. Field cooled data is shown in blue and inverse magnetic susceptibility data is shown in red. Solid black lines denote the high/low temperature fits of the inverse susceptibility data.

A Curie-Weiss fit of the high temperature section (200 – 350 K) of the inverse susceptibility taken at 1 T indicates a Weiss constant of $\theta_{CW}$ = -101.3 K and an effective moment of 4.57 $\mu_B$, in good agreement with the expected $J = 7/2$ (4.54 $\mu_B$) moment.[47] The large negative Weiss temperature indicates strong antiferromagnetic interactions. A fit of the low temperature (2-50 K) taken at 1 T indicates a smaller Weiss constant of $\theta_{CW}$ = -19.5 K and an effective moment of 3.41 $\mu_B$. To better understand the significant difference between high temperature moment ($J = 7/2$ (4.54 $\mu_B$)) and low temperature moment (3.41 $\mu_B$, unknown $J$), a literature search for analogous results was conducted. Similar measurements (also taken at low temperature and field) on structurally related ($\alpha$-NaFeO$_2$-type) single crystals of NaYbS$_2$ made in-plane (*ab* plane) report comparable values of $\theta_{CW}$ = -13.5 K and $\mu_{eff}$ = 3.2 $\mu_B$.[9] The slight deviation in value between the two structures may originate from differences between Na and Tl ionic radii, [1.02 Å and 1.50 Å, respectively] causing the structural shift from *R*-3*m* (closer layers) to *P*6$_3$/*mmc* (greater inter-layer distances). As a result of the larger non-magnetic cation (Tl) further separating the triangular lattice layers, TlYbS$_2$ more closely resembles a pure two-dimensional structure, compared to the Na analog. The report on NaYbS$_2$ describes the effective magnetic moment ($\mu_{eff}$ = 3.2 $\mu_B$) to originate from strong SOC and *ab* plane anisotropy, treating the studied material as a $J_{eff}$ = ½ system according to the two-dimensional spin ½ triangular lattice Heisenberg antiferromagnet model.[15,48] This Heisenberg model has been used to understand previously reported isostructural frustrated systems in which all magnetic ions are 120 ° from one another, which is the case for NaYbS$_2$, and thus is reasonable as a starting hypothesis for a possible magnetic model to explain the behavior of TlYbS$_2$.[31] To assess the $J_{eff}$ = ½ magnetic model, describing the proposed low temperature behavior for the Yb atoms in NaYbS$_2$, Baenitz *et al*. conducted electron-spin-resonance (ESR) measurements on single crystals and found the *g*-factor to be highly anisotropic along the two crystallographic directions ($g_{ab}$ = 3.19; $g_c$ = 0.57), an expected consequence of the triangular lattice layer.[9] Utilizing the experimentally determined $g_{ab}$-factor, the $J_{eff}$ = ½ magnetic model hypothesis, and considering equation (1), an expected moment ($\mu_{eff}$) was calculated and found to be close to their experimentally observed moment of 3.2 μ$_B$, suggesting that NaYbS$_2$ behaves as $J_{eff}$ = ½ triangular lattice Heisenberg antiferromagnet.

$$\mu_{eff} = g\sqrt{J(J+1)} \quad (1)$$

Based on the success of the analysis used for NaYbS$_2$, a similar method was employed for TlYbS$_2$ to understand the low temperature magnetic moment. Within the triangular lattice Heisenberg AFM model, the magnetization should plateau at approximately 1/3 the expected saturation magnetization, according to equation (2). In this equation, the saturation magnetization

$$m_s = J \cdot g \quad (2)$$

($m_s$) is equal to the product of the *g*-factor and the total angular momentum (*J*). By analyzing TlYbS$_2$ in the same way NaYbS$_2$ was treated using equation (1), an anisotropic *g*-factor of $g_{ab}$ = 3.94 for TlYbS$_2$ is obtained. This can be used in equation (2) to obtain a saturation magnetization saturation (1.97 μ$_B$). Taking the ratio of the moment (0.74 μ$_B$) corresponding to the inflection point observed in field dependent magnetization measurements taken at 0.42 K (**Figure 6**) with the $m_s$ discussed above yields a value of 0.37, in good agreement with the expected 0.33.

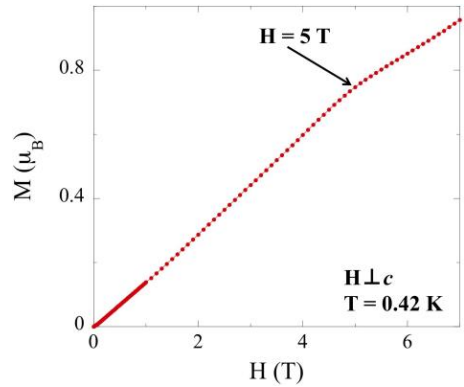

Figure 6. Field dependence of the magnetization for single crystals of TlYbS$_2$ oriented H⊥*c* over a range of 0 to 7 T at 0.42 K.

It should be noted that in the absence of a sufficiently strong magnetic field, a full plateau is not observed, but the inflection point is the onset of such a plateau. Considering the triangular lattice orientation of the Yb atoms in the *ab* plane, as shown in **Figure 4**, and the good agreement of 0.37 to the expected 0.33 as per the triangular lattice Heisenberg AFM model, it is reasonable to propose a magnetic model in which TlYbS$_2$ behaves as a $J_{eff}$ = ½ system in the low temperature (below 50 K) regime. This finding is further supported by previous reports of subtle changes in slope for inverse magnetic susceptibility plots, such as that observed in **Figure 5**, as being attributed to gradual transitions toward an isolated Kramers doublet ground state.[2]

The broad deviation from Curie-Weiss behavior at 50 K was investigated further by measuring inverse temperature dependent susceptibility perpendicular to the field at low temperatures (2 – 50 K) at fields from 0.1 T to 5 T. (**Figure 7a**) A clear deviation from linearity can be seen in the 3 T data at 4 K, with a more pronounced deviation at the same temperature for the 5 T data. Supplementary temperature dependent susceptibility measurements (**Figure 7b**) taken at low temperatures (0.4 – 2 K) and variable fields (0.2 – 7 T) reveal a gradual inversion of slope, with the 7 T data appearing as a nearly flat line. The absence of saturation in the field dependence magnetization measurements made in **Figure 6** at 7 T suggests the nearly linear susceptibility at 7 T in **Figure 7** does not correspond to true saturation.

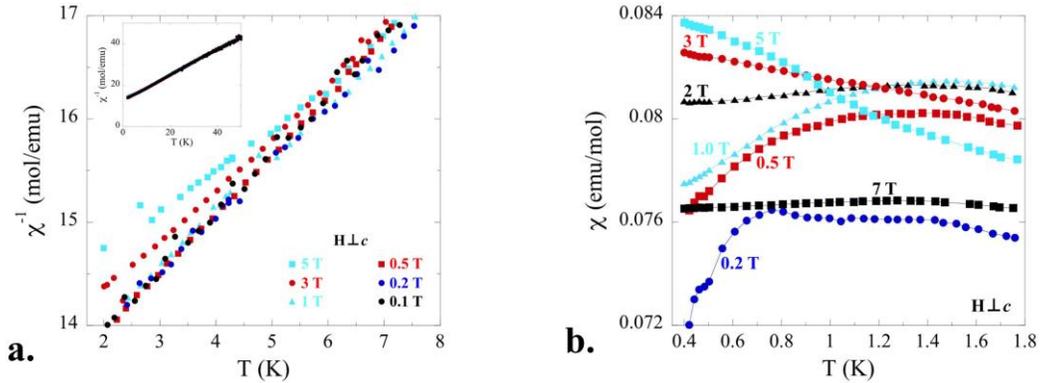

Figure 7. (a) Temperature dependence (2 – 8 K) of inverse magnetic susceptibility for single crystals of TlYbS$_2$ oriented H⊥$c$ under applied fields ranging from 0.1 T to 5 T. A wider temperature range (2 – 50 K) of the same plot is shown in the inset. (b) Temperature dependence (0.4 – 1.8 K) of susceptibility under applied fields from 0.2 T to 7 T.

Similar, albeit more subtle, features were observed in low temperature (0.4 – 1.4 K) susceptibility measurements (**Figure 8**) at variable fields (0.5 – 3 T), along the $c$ axis. The presence of such very subtle transitions and an even weaker magnetic susceptibility perpendicular to $c$ are unsurprising, as magnetic interactions are expected to reside primarily within the triangular lattice $ab$ plane. The large size of non-magnetic Tl atoms that separate the triangular lattices reasonably limit inter-layer magnetic interaction pathways. Although a complete description of this complex behavior is not possible with the current data, similar behavior has been recently reported in Os$_{0.55}$Cl$_2$ [49] and MErSe$_2$ (M = Na, K).[45] In both reports, no long-range ordering and a spin-liquid-like behavior is observed in triangular lattice frameworks similar to TlYbS$_2$, suggesting that it may also be a candidate for this exotic behavior.

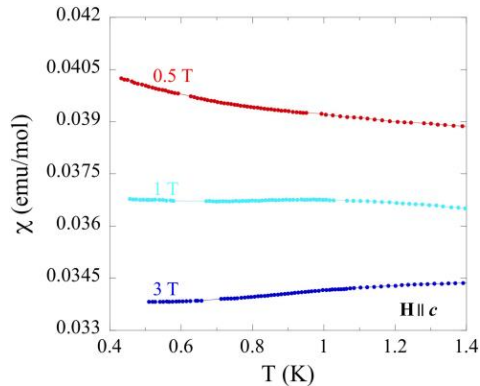

Figure 8. Temperature dependence (0.4 – 1.4 K) of magnetic susceptibility for single crystals of TlYbS$_2$ oriented H ∥ $c$ under applied fields from 0.5 T to 3 T.

## CONCLUSIONS

In summary, for the first time large single crystals of TlYbS$_2$ were prepared from a molten flux growth technique, and a reinvestigation of the nuclear structure was conducted. The single crystal structure was determined to crystallize in the hexagonal *P*6$_3$/*mmc* β-RbScO$_2$ type structure, in contrast to the previously reported trigonal *R*-3*m* α-NaFeO$_2$ structure for polycrystalline samples. Anisotropic measurements along the *ab* plane and along *c* ranging from 0.42 – 350 K and 0.1 – 7 T displayed clear short-range magnetism and metamagnetic behavior. The complete absence of long-range order and consistency of TlYbS$_2$ magnetic behavior to the two-dimensional triangular lattice Heisenberg antiferromagnet model indicates its possible candidacy as a quantum spin liquid.

## DATA AVAILABILITY

CCDS deposition number contain the 1965470 supplementary crystallographic information for this paper. These data can be obtained free of charge via www.cced.cam/ac.uk/data_request/cif, or by emailing data_request@ccdc.cam.ac.uk, or by contacting The Cambridge Crystallographic Data Center, 12 Union Road, Campbridge CB21EZ, UK; fax +44 1223 336033.

## AUTHOR CONTRIBUTIONS

Tim Ferreira was responsible for making the materials in polycrystalline and single crystal forms; additionally, he drafted this manuscript. Jie Xing was responsible for the magnetic measurements taken, and Liurukara Sanjeewa was responsible for the structural characterization. Athena Sefat oversaw the experiments and helped in finalizing this manuscript.

## FUNDING

This research, conducted at Oak Ridge National Laboratory (ORNL), is supported by the U.S. Department of Energy (DOE), Office of Science, Basic Energy Sciences (BES), Materials Sciences and Engineering Division (MSE).

## ACKNOWLEDGEMENTS

Radu Custelcean is gratefully acknowledged for permitted use of his single crystal X-ray diffractometer.

## REFERENCES

(1)  Bordelon, M. M.; Kenney, E.; Liu, C.; Hogan, T.; Posthuma, L.; Kavand, M.; Lyu, Y.; Sherwin, M.; Butch, N. P.; Brown, C.; Graf, M. J.; Balents, L.; Wilson, S. D. Field-tunable quantum disordered ground state in the triangular-lattice antiferromagnet NaYbO$_2$ *Nat. Phys.* **2019**, *15*, 1058-1064.

(2) Ranjith, K. M.; Dmytriieva, D. D.; Khim, S.; Sichelschmidt, J.; Luther, S.; Ehlers, D.; Yasuoka, H.; Wosnitza, J.; Tsirlin, A. A.; Kuhne, H.; Baenitz, M. Field-induced instability of the quantum spin liquid ground state in the Jeff = 1/2 triangular-lattice compound NaYbO$_2$ *Phys. Rev. B.* **2019**, *99*, 180401(R).

(3) Zhu, Z.; Maksimov, P. A.; White, S. R.; Chernyshev, A. L. Topography of Spin Liquids on a Triangular lattice *Phys. Rev. Lett.* **2018**, *120*, 207203.

(4) Li, Y.; Adroja, D.; Bewley, R. I.; Voneshen, D.; Tsirlin, A. A.; Gegenwart, P.; Zhang, Q. Crystalline Electric-Field Randomness in the Triangular Lattice Spin-Liquid YbMgGaO$_4$ *Phys. Rev. Lett.* **2017**, *118*, 107202.

(5) Shen, Y.; Dong Li, Y.; Wo, H.; Li, Y.; Shen, S.; Pan, B.; Wang, Q.; Walker, H. C.; Steffens, P.; Boehm, M.; Hao, Y.; Quintero-Castro, D. L.; Harriger, L. W.; Frontzek, M. D.; Hao, L.; Meng, S.; Zhang, Q.; Chen, G.; Zhao, J. Evidence for a spinon Fermi surface in a triangular-lattice quantum-spin-liquid candidate *Nature* **2016**, *540*, 559-562.

(6) Lee, P. A.; Nagaosa, N.; Wen, X.-G. Doping a Mott insulator: Physics of high-temperature superconductivity *Rev. Mod. Phys.* **2006**, *78*, 17.

(7) Chubokov, A. V.; Golosov, D. I. Quantum theory of an antiferromagnet on a triangular lattice in a magnetic field *J. Phys.: Condens. Matter* **1991**, *3*, 69.

(8) Savary, L.; Balents, L. Quantum spin liquids: A review *Rep. Prog. Phys.* **2016**, *80*, 016502.

(9) Baenitz, M.; Schlender, P.; Sichelschmidt, J.; Onykiienko, Y. A.; Zangeneh, Z.; Ranjith, K. M.; Sarkar, R.; Hozoi, L.; Walker, H. C.; Orain, J.-C.; Yasuoka, H.; Brink, J. V. D.; Klauss, H. H.; Inosov, D. S.; Doert, T. NaYbS2: A planar spin-1/2 triangular-lattice magnet and putative spin liquid *Phys. Rev. B.* **2018**, *98*, 220409(R).

(10) Li, Y.; Adroja, D.; Voneshen, D.; Bewley, R. I.; Zhang, Q.; Tsirlin, A. A.; Gegenwart, P. Nearest-neighbour resonating valence bonds in YbMgGaO$_4$ *Nat. Commun.* **2017**, *8*, 15814.

(11) Paddison, J. A. M.; Daum, M.; Dun, Z.; Ehlers, G.; Liu, Y.; Stone, M. B.; Zhou, H.; Mourigal, M. Continuous excitations of the triangular-lattice quantum spin liquid YbMgGaO4 *Nat. Phys.* **2017**, *13*, 117.

(12) Dun, Z. L.; Trinh, J.; Li, K.; Lee, M.; Chen, K. W.; Baumbach, R.; Hu, Y. F.; Wang, Y. X.; Choi, E. S.; Shastry, B. S.; Ramirez, A. P.; Zhou, H. D. Magnetic Ground States of the Rare-Earth Tripod Kagome Lattice Mg$_2$RE$_2$Sb$_3$O$_{14}$ (RE = Gd, Dy, Er) *Phys. Rev. Lett.* **2016**, *116*, 157201.

(13) Xu, Y.; Zhang, J.; Li, Y. S.; Hong, X. C.; Zhang, Q. M.; LI, S. Y. Absence of Magnetic Thermal Conductivity in the Quantum Spin-liquid Candidate YbMgGaO$_4$ *Phys. Rev. Lett.* **2016**, *117*, 267202.

(14) Starykh, O. A. Unusual ordered phases of highly frustrated magnetism: a review. *Rep. Prog. Phys.* **2015**, *78*, 052502.

(15) Anderson, P. W. Resonating valence bonds: A new kind of insulator *Mater. Res. Bull.* **1973**, *8*, 153.

(16) Hu, W.-J.; Gong, S.-S.; Zhu, W.; Sheng, D. N. Competing spin-liquid states in the spin-1/2 Heisenberg model on the triangular lattice *Phys. Rev. B.* **2015**, *92*, 140403(R).

(17) Balents, L. Spin liquids in frustrated magnets *Nature* **2010**, *464*, 199-208.

(18) Yamamoto, D.; Marmorini, G.; Danshita, I. Microscopic Model Calculatons for the Magnetization Process of Layered Triangular-Lattice Quantum Antiferromagnets *Phys. Rev. Lett.* **2015**, *114*, 027201.

(19) Shirata, Y.; Tanaka, H.; Matsuo, A.; Kindo, K. Experimental realization of a spin-1/2 triangular-lattice Heisenberg antiferromagnet *Phys. Rev. Lett.* **2012**, *108*, 057205.

(20) Zhou, H. D.; Choi, E. S.; Li, G.; Balicas, L.; Wiebe, C. R.; Qiu, Y.; Copley, J. R. D.; Gardner, J. S. Spin Liquid State in the S = 1/2 Triangular Lattice $Ba_3CuSb_2O_9$ *Phys. Rev. Lett.* **2011**, *106*, 147204.

(21) Yamashita, M.; Nakkata, N.; Senshu, Y.; Nagata, M.; Yamamoto, H. M.; Kato, R.; Shibauchi, T.; Matsuda, Y. Highly Mobile Gapless Excitations in a Two-Dimensional Candidate Quantum Spin Liquid *Science* **2010**, *328*, 1246.

(22) Yoshida, M.; Takigawa, M.; Yoshida, H.; Okamoto, Y.; Hiroi, Z. Phase Diagram and Spin Dynamics in Volborthite with a Distorted Kagome Lattice *Phys. Rev. Lett.* **2009**, *103*, 077207.

(23) Helton, J. S.; Matan, K.; Shores, M. P.; Nytko, E. A.; Bartlett, B. M.; Yoshida, Y.; Takano, Y.; Suslov, A.; Qui, Y.; Chung, J.-H.; Nocera, D. G.; Lee, Y. S. Spin Dynamics of the Spin-1/2 Kagome Lattice Antiferromagnet $ZnCu_3(OH)_6Cl_2$ *Phys. Rev. Lett.* **2007**, *98*, 107204.

(24) Lee, S.-S.; Lee, P. A. U(1) Gauge Theory of the Hubbard Model: Spin Liquid States and Possible Application to k-$(BEDT-TTF)_2Cu_2(CN)_3$ *Phys. Rev. Lett.* **2005**, *95*, 036403.

(25) Onoda, S.; Tanaka, Y. Quantum Melting of Spin Ice: Emergent Cooperative Quadrupole and Chirality *Phys. Rev. Lett.* **2010**, *105*, 047201.

(26) Gardner, J. S.; Gingras, M. J. P.; Greedan, J. E. Magnetic pyrochlore oxides *Rev. Mod. Phys.* **2010**, *82*, 53.

(27) Curnoe, S. H. Structural disortion and the spin liquid state in $Tb_2Ti_2O_7$ *Phys. Rev. B.* **2008**, *78*, 094418.

(28) Okamoto, Y.; Nohara, M.; Aruga-Katori, H.; Takagi, H. Spin-Liquid State in the S = 1/2 Hyperkagome Antiferromagnet $Na_4Ir_3O_8$ *Phys. Rev. Lett.* **2007**, *99*, 137207.

(29) Applegate, R.; Hayre, N. R.; Singh, R. R. P.; Lin, T.; Day, A. G. R.; Gingras, M. J. P. Vindication of $Yb_2Ti_2O_7$ as a Model Exchange Quantum Spin Ice *Phys. Rev. Lett.* **2012**, *109*, 097205.

(30) Laconis, J.; Liu, C.; Halaz, G. B.; Balents, L. Spin liquid versus spin orbit coupling on the triangular lattice *SciPost Phys.* **2018**, *4*, 003.

(31) Liu, W.; Zhang, Z.; Ji, J.; Liu, Y.; Li, J.; Wang, X.; Lei, H.; Chen, G.; Zhang, Q. Rare-Earth Calcogenides: A Large Family of Triangular Lattice Spin Liquid Candidates *Chin. Phys. Lett.* **2018**, *35*, 117501.

(32) Lu, L.; Song, M.; Liu, W.; Reyes, A. P.; Kuhns, P.; Lee, H. O.; Fisher, I. R.; Mitrovic, V. F. Magnetism and local symmetry breaking in a Mott insulator with strong spin orbit interactions *Nat. Commun.* **2017**, *8*, 14407.

(33) Li, Y.-D.; Wang, X.; Chen, G. Anisotropic spin model of strong spin-orbit-coupled triangular antiferromagnets *Phys. Rev. B.* **2016**, *94*, 035107.

(34) Li, Y.; Chen, G.; Tong, W.; Pi, L.; Liu, J.; Yang, Z.; Wang, X.; Zhang, Q. Rare-Earth Triangular Lattice Spin Liquid: A Single-Crystal Study of $YbMgGaO_4$ *Phys. Rev. Lett.* **2015**, *115*, 167203.

(35) Zhu, Z.; Maksimov, P. A.; White, S. R.; Cheryshev, A. L. Disorder-Induced Mimicry of a Spin Liquid in $YbMgGaO_4$ *Phys. Rev. Lett.* **2017**, *119*, 157201.

(36) Duczmal, M.; Pawlak, L. Magnetic properties of $TlLnS_2$ compounds (Ln = Nd, Gd, Dy, Er, and Yb) *J. Alloys Compd.* **1994**, *209*, 271-274.

(37) (2015), B. APEX3, SAINT+, and SADABS. Madison, WI: Bruker AXS Inc.

(38) Sheldrick, G. M. A Short History of SHELX *Acta Crystallogr., Sect. A: Found. Crystallogr.* **2008**, *64*, 112-122.

(39) Stowe, K. The Crystal Strctures of ErSeI and $NaErSe_2$ *Z. anog. all. Chem.* **1997**, *623*, 1639-

1643.

(40) Bugaris, D. E.; zur Loye, H.-C. Materials Discovery by Flux Crystal Growth: Quaternary and Higher Order Oxides *Angew. Chem., Int. Ed.* **2012**, *51*, 3780-3811.

(41) Usman, M.; Smith, M. D.; Morrison, G.; Klepov, V. V.; Zhang, W.; Halasyamani, P. S.; zur Loye, H.-C. Molten Alkali Halide Flux Growth of an Extensive Family of Noncentrosymmetric Rare Earth Sulfides: Structure and Magnetic Optical (SHG) Properties *Inorg. Chem.* **2019**, *58*, 8541-8550.

(42) Usman, M.; Morrison, G.; zur Loye, H.-C. $La_2USe_3S_2$: A Serendipiously Grown Lanthanide/Actinide Chalcogenide from a Eutectic Halide Flux *J. Chem. Cryst.* **2019**, *49*, 169-173.

(43) Tsujimoto, Y.; Juillerat, C. A.; Zhang, W.; Fujii, K.; Yashima, M.; Halasyamani, P. S.; zur Loye, H.-C. Function of Tetrahedral $ZnS_3O$ Building Blocks in the Formation of $SrZn2S2O$: A Phase Matchable Polar Oxysulfide with a Large Second Harmonic Generation Response *Chem. Mater.* **2018**, *30*, 6486-6493.

(44) Klepov, V. V.; zur Loye, H.-C. Complex Topologies from Simple Building Blocks: Uranium(IV) Thiophosphates *Inorg. Chem.* **2018**, *57*, 11175-11183.

(45) Xing, J.; Sanjeewa, L. D.; Kim, J.; Meier, W. R.; May, A. F.; Zheng, Q.; Custelcean, R.; Stewart, G. R.; Sefat, A. S. Synthesis, magnetization and heat capacity of triangular lattice materials $NaErSe_2$ and $KErSe_2$ *Submitted.* **2019**.

(46) Xing, J.; Sanjeewa, K. D.; Kim, J.; Stewart, G. R.; Du, M.-H.; Reboredo, F.A.; Custelcean, R.; Sefat, A. S. Crystal Synthesis and Frustrated Magnetism in Triangular Lattice $CsRESe2$ (RE = La - Lu): Quantum Spin Liquid Candidates $CsCeSe_2$ and $CsYbSe_2$ *Submitted*. **2019**.

(47) Hashimoto, Y.; Wakeshima, M.; Hinatsu, Y. Magnetic properties of ternary sodium oxides $NaLnO_2$ (Ln - rare earths) *J. Solid State Chem.* **2003**, *176*, 266.

(48) Huse, D. A.; Elser, V. Simple Variational Wave Functions for Two-Dimensional Heisenberg Spin-1/2 Antiferromagnets *Phys. Rev. Lett.* **1988**, *60*, 2531.

(49) McGuire, M. A.; Zheng, Q.; Yan, J.; Sales, B. C. Chemical disorder and spin-liquid-like magnetism in the van der Waals layered 5d transition metal halide $Os_{0.55}Cl_2$ *Phys. Rev. Lett.* **2019**, *99*, 214402.

Table 1. Crystallographic data for TlYbS$_2$ determined by single crystal X-ray diffraction

| Empirical formula | TlYbS$_2$ |
|---|---|
| Formula weight (g/mol) | 441.53 |
| $T$, K | 273 |
| Crystal habitat | red plates |
| Crystal dimensions, mm | 0.060 x 0.020 x 0.020 |
| Crystal system | hexagonal |
| Space group | $P6_3/mmc$ (No.194) |
| $a$, Å | 3.9454(8) |
| $c$, Å | 15.277(7) |
| Volume, Å$^3$ | 205.94(12) |
| $Z$ | 2 |
| $D$ (calc), g/cm$^3$) | 7.120 |
| $\mu$ (Mo K$\alpha$), mm$^{-1}$ | 62.409 |
| $F$(000) | 366 |
| $T$max, $T$min | 0.2092-1.0000 |
| $\theta$ range | 2.67 - 30.63 |
| Reflections collected | 1816 |
| Data/restraints/parameters | 102/0/9 |
| Final $R$ [$I> 2\sigma(I)$] $R_1$, $R_{w2}$ | 0.0366/0.0797 |
| Final $R$ (all data) $R_1$, $R_{w2}$ | 0.0402/0.0825 |
| GoF | 1.069 |
| Largest diff. peak/hole, e/ Å$^3$ | 1.831/-1.820 |

Table 2. Atomic coordinates and equivalent isotropic displacement parameters (Å$^2$) for TlYbS$_2$

| Atom | Wyckoff | x | y | z | $U_{eq}$ (Å$^2$) |
|---|---|---|---|---|---|
| Tl | 2c | 0.33333 | 0.66667 | 0.25000 | 0.0471(10) |
| Yb | 2b | 0 | 0 | 0 | 0.0268(8) |
| S | 4f | 0.33333 | 0.66667 | 0.5965(6) | 0.0278(19) |

Table 3. Bond distances (Å) and angles (°) for TlYbS$_2$

|  | Bond distances (Å) |  | Bond Angles (°) |
| --- | --- | --- | --- |
| Yb–S (6x) | 2.713(5) | Yb–S–Yb | 93.3(2) |
| Yb–Yb | 3.9454(8) | Yb–Yb–Yb | 60.0 |

Table 4. Energy dispersive spectroscopy data

| Element | Atomic % |
|---------|----------|
| Tl | 25.99 |
| Yb | 25.10 |
| S | 48.91 |